\documentclass[a4paper,10pt,twocolumn]{article}
\usepackage[utf8]{inputenc}
\usepackage{multicol,graphicx}
\usepackage{mathptmx}
\usepackage[T1]{fontenc}
\usepackage{textcomp}
\usepackage{hyperref}
\usepackage{url}
\usepackage{subfig}

\usepackage[T1]{fontenc}
\usepackage[english]{babel}

\usepackage[backend=bibtex,style=ieee,doi=false,isbn=false,url=true,maxnames=6,citestyle=numeric-comp,giveninits=true]{biblatex}
\fontfamily{ptm}\selectfont

\bibliography{references.bib}

\usepackage[font=small]{caption}
\usepackage{amssymb}
\usepackage{amsmath}
\usepackage{mathrsfs}

\newcommand{\mysection}[1]{\vspace{0.4cm} \uppercase{#1} \vspace{0.4cm}}
\newcommand{\mysubsection}[1]{\hspace{10pt}\textit{#1:}}

\begin{document}
	
\setlength{\textfloatsep}{10pt plus 1.0pt minus 2.0pt}	
\setlength{\columnsep}{1cm}


\twocolumn[%
\begin{@twocolumnfalse}
\begin{center}
	{\fontsize{14}{18}\selectfont
        \textbf{\uppercase{Enhancing Computational Efficiency of Motor Imagery BCI classification with Block-Toeplitz Augmented Covariance Matrices and Siegel metric}}\\}
    \begin{large}
        \vspace{0.6cm}
        Igor Carrara\textsuperscript{1}, Theodore Papadopoulo\textsuperscript{1}\\
        \vspace{0.6cm}
        \textsuperscript{1}Université Côte d'Azur, INRIA, Cronos Team, France\\
        \vspace{0.5cm}
        E-mail: igor.carrara@inria.fr, theodore.papadopoulo@inria.fr
        \vspace{0.4cm}
    \end{large}
\end{center}	
\end{@twocolumnfalse}%
]%

ABSTRACT: Electroencephalographic signals are represented as multidimensional datasets. We introduce an enhancement to the augmented covariance method (ACM), exploiting more thoroughly its mathematical properties, in order to improve motor imagery classification.
Standard ACM emerges as a combination of phase space reconstruction of dynamical systems and of Riemannian geometry. Indeed, it is based on the construction of a Symmetric Positive Definite matrix to improve classification. But this matrix also has a Block-Toeplitz structure that was previously ignored. This work treats such matrices in the real manifold to which they belong: the set of Block-Toeplitz SPD matrices. After some manipulation, this set is can be seen as the product of an SPD manifold and a Siegel Disk Space.
The proposed methodology was tested using the MOABB framework with a within-session evaluation procedure. It achieves a similar classification performance to ACM, which is typically better than -- or at worse comparable to -- state-of-the-art methods. But, it also improves consequently the computational efficiency over ACM, making it even more suitable for real time experiments.


\mysection{Introduction}

In electroencephalography (EEG) based Brain Computer Interfaces (BCI), state-of-the-art algorithms are often built on Riemannian distance based classification algorithms~\cite{barachant-bonnet-etal:10}. The basic idea underlying these methods is to treat the spatial covariance matrix (SCM), extracted from the EEG signal, as an element of the Riemannian manifold of Symmetric Positive Definite (SPD) matrices~\cite{forstner-moonen:03}.

A recent extension of this work was obtained by using the Augmented Covariance Method (ACM)~\cite{carrara-papadopoulo:23}. ACM relies on the concept of phase space reconstruction of dynamical systems to create an "ACM matrix" (also called ACM) that contains not only an average spatial representation of the signal but also a representation of its evolution in time. Consequently, the amount of information contained in this ACM matrix is increased w.r.t. the standard spatial covariance. As the ACM matrix also turns out to be an SPD matrix, it can be classified using the same Riemannian framework that was so successful for SCMs. However, it also possesses a structural property of being Block-Toeplitz, that is, a block matrix with constant blocks over all diagonals. Recently, an approach has been proposed to better deal with such Block-Toeplitz SPD matrices~\cite{jeuris-vandebril:16}, with applications in diverse fields such as audio processing or radar signal analysis~\cite{cabanes:22}.

The idea of this research is thus to endow the smooth manifold of Block-Toeplitz SPD matrices with a Riemannian metric, thus allowing the ACM matrix to be treated within its true manifold membership. It is actually possible to treat the Block-Toeplitz SPD matrix manifold as the product of an SPD manifold and a Siegel Disk Space, after applying an appropriate conversion of the blocks of the ACM matrix into the Verblusky coefficients~\cite{dette-wagener:10}.

This approach provides a new -- more specific -- metric to use for BCI classification algorithms. The strength of the approach lies in its ability to deconstruct the manifold into its constituent elements: the Symmetric Positive Definite (SPD) manifold and the Siegel Disk Space. By discerningly analyzing each component within its respective geometrical domain, this method significantly alleviates the computational demand traditionally associated with the ACM methodology. The resulting algorithm achieves performance that is, at worst, comparable with state-of-the-art BCI algorithms, and often provides quite better results (on par with those of ACM). However, it distinguishes itself by achieving this at substantial reduction in computational costs and carbon footprint compared with standard ACM.

The new -- Siegel metric based -- pipeline was tested and validated against several state-of-the-art algorithms (Machine Learning (ML) and Deep Learning (DL)) on several datasets for motor imagery (MI) classification using several subjects and on a right versus left hand task with the \href{http://moabb.neurotechx.com/docs/index.html}{MOABB} framework~\cite{aristimunha-carrara-etal:23}, and a within-session evaluation procedure. 


\mysection{Materials and Methods}

The EEG signal is represented as a multivariate time series $\mathbf{X} \in \mathbb{R}^{d \times T}$, where $T$ represents the total number of sampled data points, and $d$ indicates the number of electrodes used in the EEG recording. Since this paper focuses on MI task, we split the EEG signals into smaller sections known as epochs, each representing a snapshot of brain activity during various tasks or mental states. The core aim of our research is to develop a method that can accurately identify the specific task or mental state associated with these EEG epochs.

The space of SPD matrices is composed by square real symmetric matrices that are positive definite, and this space form a smooth manifold that can be equipped with a Riemannian metric~\cite{barachant-bonnet-etal:10}. This space is defined as
\begin{equation}
    \mathbb{SPD}_d=\{\mathbf{M} \in \mathbb{R}^{d \times d}\; |\; x^T\mathbf{M}x>0 \; \; \forall x \in \mathbb{R}^d \symbol{92} \{0\}\}
\end{equation}
ACM~\cite{carrara-papadopoulo:23} extends this methodology by combining it with the phase space reconstruction (PSR) approach that is grounded in the Takens theorem~\cite{takens:81}. The ACM matrix thus obtained contains spatial and temporal information of the signal and remains an SPD matrix that can be classified using the same Riemannian metric that was so successful for SCMs. This enrichment with temporal features of the information extracted from the signal allows for an improvement of classification performance.

The idea of using Takens theorem is based on the idea that time series obtained from experimental observations that capture only a fraction of the complex dynamics of the underlying system, can nonetheless be utilized to reconstruct the system's full dynamical behavior. This is achieved using a uniform embedding procedure: consider a time series ${s(n)}$ created thought a measurement process, the PSR technique generates a point $\mathbf{s}_E(n)$ of a $D$-dimensional space constructed as
\begin{equation}
    \mathbf{s}_E(n) = [s(n), s(n - \tau), ..., s(n - (D - 1)\tau)]^T
\end{equation}
where $\tau$ is a positive integer called the embedding delay and $D$ is the embedding dimension. $\mathbf{s}_E(n) \in \mathbb{R}^{D}$ is an uniform embedding of the original phase space. 

\begin{figure*}[!ht]
\centering
\includegraphics[width=\linewidth]{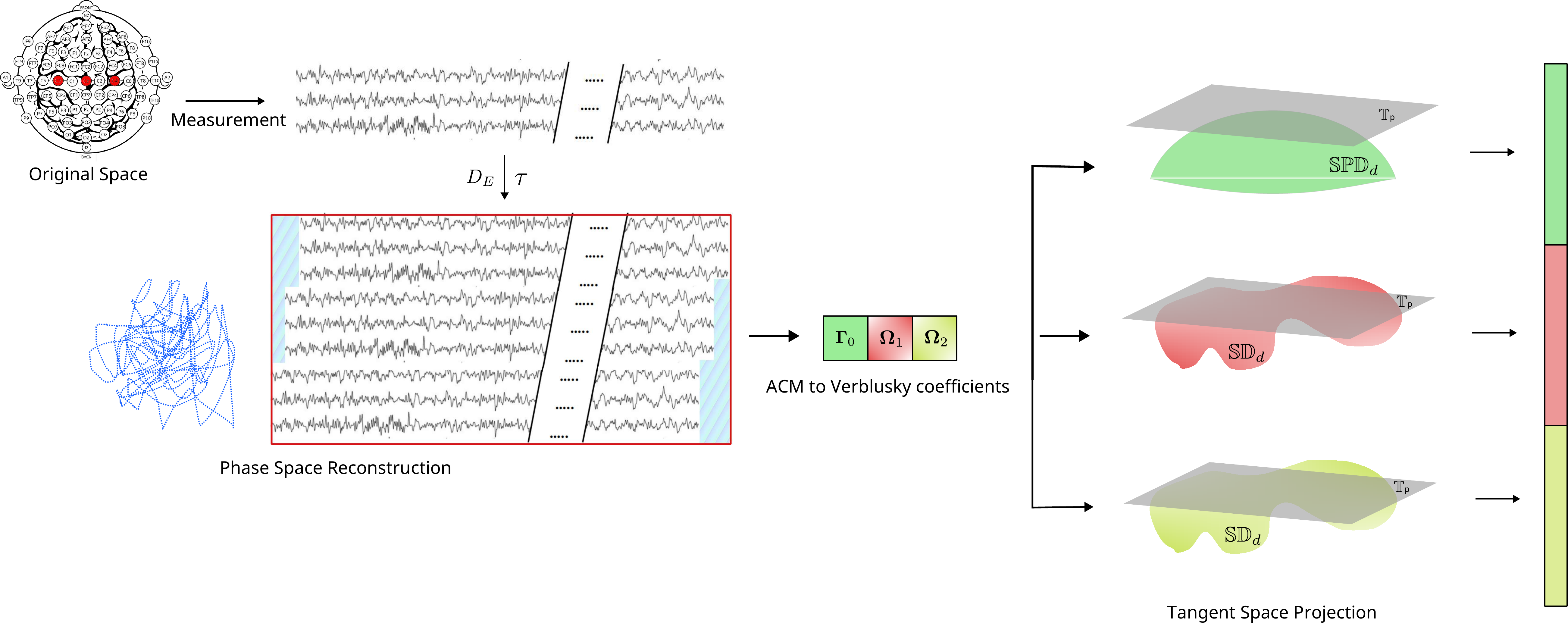}
\caption{Schematic illustration of the BT-ACM + TG + SVM methodology. The presented example uses only 3 electrodes (in red on the top left plot). The measurement process of the original dynamic system is thus a 3-dimensional time series. The process then begins with the extraction of epoch signal representing left and right hand tasks. We then use the phase space reconstruction process to obtain a dynamic system equivalent to the original one (selection of hyper-parameters are made via grid search using the nested approach). In this figure, we see an embedding corresponding to $p=3$ and $\tau=10$). The BT-ACM matrix is computed as the autocovariance of this high-dimensional time series. Subsequently the main blocks are converted in Verblunsky coefficients. Then, each component is mapped to the tangent space using the appropriate Riemannian manifold computations and vectorized. The final step is the application of an SVM-based classification algorithm.
}
\label{fig:Total}
\end{figure*}

The ACM matrix (see Fig.~\ref{fig:Total}) is obtained by expanding the original EEG signal using the PSR approach with an embedding dimension $p$ to get a new $dp \times T$ time series, parameterized by the fixed delay $\tau$. The Augmented Covariance Matrix $\Gamma_{aug}$ is defined as the autocovariance matrix of this new time series:
\begin{equation}
    \Gamma_{Aug} = \left[\begin{array}{cccc}
\Gamma_{0} & \Gamma_{-1} & \Gamma_{-2} & \cdots \\
\Gamma_{1} & \Gamma_{0} & \Gamma_{-1} & \cdots \\
\Gamma_{2} & \Gamma_{1} & \Gamma_{0} & \cdots \\
\vdots & \vdots & \vdots & \ddots \\
\Gamma_{p-1} & \Gamma_{p-2} & \Gamma_{p-3} & \cdots
\end{array}\right]\; ,
\end{equation}
where $\Gamma_{0}$ is the standard spatial covariance matrix and $\Gamma_{i}$ is the lagged covariance matrix of the original signal with a delay of $i\,\tau$ and $\Gamma_{-i} = \Gamma_{i}^T$.
As an autocovariance matrix, $\Gamma_{aug}$ is symmetric and positive by construction. If not definite,  it can be regularized~\cite{chen-wiesel-etal:10}, so that we consider it as SPD in the remainder of this article. But, the ACM matrix also has a specific Block-Toeplitz structure, with blocks of dimension $d\times d$~\cite{cabanes:22}. More formally, $\Gamma_{Aug}$ belongs to the space $\mathbb{B}_{d \times p}$ of Block-Toeplitz and SPD matrices i.e., SPD matrices of size $dp\times dp$ with constant blocks of size $d\times d$ along all diagonals. This opens up new possibilities for enhancing the ACM formulation by mapping the ACM matrix to the most suitable geometric space that fully captures both its Block-Toeplitz and SPD natures.

The blocks of the matrix $\Gamma_{Aug}$, have been demonstrated to belong to a specific mathematical space~\cite{jeuris-vandebril:16}
\begin{equation}
    \Gamma_{i} \in \mathbb{D}_d \; \; \; \; \mathbb{D}_d = \{\mathbf{M} \in \mathbb{C}^{d \times d} \; | \; \mathbf{I}- \mathbf{M}\Bar{\mathbf{M}}>0 \}
\end{equation}
with $\Bar{\mathbf{M}} = \mathbf{J}\mathbf{M}^H\mathbf{J}$ where $\mathbf{J}$ denotes the anti-diagonal matrix and $^H$ is the conjugate transpose operator\footnote{$A > B$ when $A - B$ is a positive definite matrix.}.
This space has no known Riemannian structure but, by implementing a minor adjustment to the coefficients, it is possible to ensure their belonging within the domain of the Siegel disk~\cite{dette-wagener:10, fritzsche-kirstein:87}, defined as 
\begin{equation}
    \Omega_{i} \in \mathbb{SD}_d \; \; \; \; \mathbb{SD}_d = \{\mathbf{M} \in \mathbb{C}^{d \times d} \; | \; \mathbf{I}- \mathbf{M}\mathbf{M}^H>0 \}\;.
\end{equation}
The coefficients that have undergone such modification are also known as Verblunsky coefficients~\cite{dette-wagener:10}.
 
The transformation allow the following conversion, 
\begin{equation}
    \begin{aligned}
    \mathbb{B}_{d \times p} & \rightarrow \mathbb{SPD}_d \times \mathbb{SD}_d^{p-1} \\
    \Gamma_{Aug} & \mapsto \left(\Gamma_{0}, \Omega_1, \ldots, \Omega_{p-1}\right) .
\end{aligned}
\end{equation}
Consider the initial matrix $\Gamma_{Aug}$ decomposed in its constituent blocks $(\Gamma_{0}, ..., \Gamma_{p-1})$. The initialization of the recursive transformation is set to $\mathbf{P}_0 = \Gamma_0$. The subsequent coefficients are computed with
\begin{equation}
    \Omega_{l+1} = \mathbf{L}_l^{-1/2}(\mathbf{R}_{l+1}-\mathbf{M}_l)\mathbf{K}_l^{-1/2} \;,
    \label{eq:Verblunsky}
\end{equation} 
with $l = 0, ..., p-1$ and
\begin{eqnarray*}
    \mathbf{L}_l & = &\mathbf{P}_0 - (\Gamma_{1}, ..., \Gamma_{l})\Tilde{\Gamma}^{-1}_{l-1}(\Gamma_{1}, ..., \Gamma_{l})^H \\
    \mathbf{K}_l & = &\mathbf{P}_0 - (\Gamma_{1}^H, ..., \Gamma_{l}^H)\Tilde{\Gamma}^{-1}_{l-1}(\Gamma_{1}^H, ..., \Gamma_{l}^H)^H \\
    \mathbf{M}_l & = &(\Gamma_{1}, ..., \Gamma_{l})\Tilde{\Gamma}^{-1}_{l-1}(\Gamma_{1}^H, ..., \Gamma_{l}^H)^H
\end{eqnarray*}
where $\Tilde{\Gamma}_{l-1}$ denotes the sub-matrix of $\Gamma_{Aug}$ obtained by keeping only its first $l-1$ rows and columns. This transformation operates recursively, enabling the foundational blocks of the $\Gamma_{Aug}$ matrix to be transformed into square matrices that are positioned within the domain of the Siegel Disk.

The smooth manifold of $\mathbb{B}_{d \times p}$ is thus identified as a Kähler manifold~\cite{jeuris-vandebril:16}, on which is possible to define a Kähler potential $\Phi$~\cite{ballmann:06, yang:11}, computed as:
\begin{equation}
    \Phi(\Gamma_{Aug}) = - log(det(\Gamma_{Aug}))-log(\pi e)
\end{equation}
After applying some decomposition properties of the determinant of $\Gamma_{Aug}$, it is possible to compute the metric of the manifold simply as the Hessian matrix of the Kähler potential
\begin{equation}
    \begin{aligned}
d s^2= & p \operatorname{trace}\left(\mathbf{P}_0^{-1} d \mathbf{P}_0 \mathbf{P}_0^{-1} d \mathbf{P}_0\right) \\
& +\sum_{l=1}^{p-1}(p-l) \operatorname{trace}\left(\left(\mathbf{I}-\Omega_l \Omega_l^H\right)^{-1} d \Omega_l\left(\mathbf{I}-\Omega_l^H \Omega_l\right)^{-1} d \Omega_l^H\right)
\end{aligned}
\label{Metric}
\end{equation}
The first term of Equation~(\ref{Metric}) is identifiable as the metric for the $\mathbb{SPD}_d$ space. The other term represents the metric for the Siegel disk space $\mathbb{SD}_d$ repeated $p-1$ times, i.e. the space of Block-Toeplitz SPD matrices is equipped with a product Riemannian metric over $\mathbb{SPD}_d \times \mathbb{SD}_d^{p-1}$.

The resulting algorithm (depicted in Fig.~\ref{fig:Total}) using this new metric using SVM on the tangent space is called BT-ACM+TS+SVM: Block-Toeplitz Augmented Covariance Matrix (BT-ACM) with Tangent Space projection (TS) and SVM classifier. 

Using fixed hyper-parameters $p$ and $\tau$, the PSR approach expands the signal ($p$ and $\tau$ will subsequently be carefully chosen using a grid-search procedure). The spatial autocovariance matrix of the expanded signal is obtained using regularization through the Oracle Approximating Shrinkage Estimator (OAS)~\cite{ chen-wiesel-etal:10}. After the Verblunsky transformation, each component is mapped to the Tangent space using the Logarithmic map of each specific manifold. The final classification step is obtained with a Support Vector Machine (SVM) algorithm.

\mysubsection{Dataset and Evaluation procedure}
To validate the proposed methodology, we use open accessible datasets available from MOABB~\cite{aristimunha-carrara-etal:23}. We selected 3 datasets with a total of 70 subjects. All the information regarding the considered datasets are presented in Tab.~\ref{table:dataset}.

\begin{table}[!ht]
\caption{Dataset considered during this study.}
\resizebox{\linewidth}{!}{\begin{tabular}{c|c|c|c|c|c}
\text { Dataset }                              & \text { subjects } & \text { channels } & \text {sampling rate } & \text { trials/class } & \text{Epoch (s)} \\
\hline \text {BNCI2014001~\cite{tangermann-muller-etal:12}} & 9     & 22                 & 250 Hz                           & 144                    & [2, 6] \\                   
\hline \text {BNCI2014004~\cite{leeb-lee-etal:07}} & 9              & 3                  & 250 Hz                           & 360                    & [3, 7.5]\\     
\hline \text {Cho2017~\cite{cho-ahn-etal:17}}             & 52      & 64                 & 512 Hz                           & 100                    & [0, 3] \\
\hline 
\end{tabular}}
\label{table:dataset}
\end{table}

The duration of each epoch within our study is intentionally aligned with the task conditions' length, which is subject to variation across the datasets employed. On each dataset, we applied a standard band pass filter procedure for the Motor Imagery task, in the frequency range of 8 to 32 Hz.

We use a Within-Session (WS) evaluation procedure as provided in MOABB. This means that our analysis works on each session separately. The implementation is based on a Nested Cross-Validation methodology~\cite{cawley-talbot:10}, structured with an outer loop of 5-Fold Cross validation and a inner one composed by a 3-fold Cross Validation. We use statistical tests provided by MOABB to confront the different pipelines, i.e., based on a t-test~\cite{student:08} for datasets with less than 20 subjects, or a Wilcoxon non-parametric signed-rank test~\cite{wilcoxon:92} otherwise.

The state-of-the-art pipelines used in this research contain both Machine Learning (ML) and Deep Learning (DL) methods. Detail of the pipelines are listed in Tab.~\ref{table:Pipeline}.

\begin{table*}[!ht]
\caption{Pipelines considered in this study are organized into two distinct sections within the table: the first part is dedicated to the traditional classical ML pipelines, while the second part focuses on DL pipelines for MI.} 
\label{table:Pipeline} 
\resizebox{\linewidth}{!}{\begin{tabular}{|l|l|l|}
\hline
\textbf{Pipeline} & \textbf{Feature Extraction} & \textbf{Classifier} \\
\hline 
CSP + LDA~\cite{lotte-guan:10b} & Common Spatial Patterns (CSP) with OAS covariance estimator & Optimized Shrinkage LDA \\
MDM~\cite{barachant-bonnet-etal:10} & Spatial Covariance using OAS & Mean Distance to Mean (MDM) \\
FgMDM~\cite{barachant-bonnet-etal:10} & Spatial Covariance using OAS & Minimum Distance to Mean with geodesic filtering (FgMDM) \\
TS + EL~\cite{corsi-chevallier-etal:22} & Spatial Covariance using OAS mapped to TS & Optimized Elastic Network (EL) \\
TS + SVM~\cite{barachant-bonnet-etal:10} & Spatial Covariance using OAS mapped to TS & Optimized SVM \\
ACM + TS + SVM~\cite{carrara-papadopoulo:23} & ACM with Sample Covariance Estimator mapped to TS & Optimized SVM \\
BT-ACM + TS + SVM (Proposed) & BT-ACM with Sample Covariance using OAS mapped to each respectively TS & SVM \\
\hline
ShallowConvNet~\cite{schirrmeister-springenberg-etal:17} & Standardized and resample EEG signal at 250Hz & Convolutional Neural Network (CNN) \\
DeepConvNet~\cite{schirrmeister-springenberg-etal:17} & Standardized and resample EEG signal at 250Hz & CNN \\
EEGNet 8 2~\cite{lawhern-solon-etal:18} & Standardized and resample EEG signal at 128Hz & CNN with architecture EEGNet \\
\hline
\end{tabular}}
\end{table*}

For DL pipelines, we used a standardization step that normalizes every channel to have a zero mean and unit standard deviation. Additionally, we employed a re-sampling procedure to ensure that each architecture integrates a temporal filter aligned with the state-of-the-art techniques' implementations. This procedure was added in order align to the state-of-the-art implementation and avoid the need of redoing hyper-parameter tuning. The DL pipelines are using a Sparse Categorical Cross-Entropy loss function and a standard Adam optimizer using 300 epochs and a batch size of 64. To avoid overfitting, we used an early stopping procedure with a patience parameter of 75.


\mysection{Results}

In this section, we describe the results obtained for the Right vs Left hand classification task.

\begin{table}[!ht]
\caption{Performance (AUC) of Right hand vs Left hand classification. The table contains the results over all subjects (average plus or minus standard deviation).}
\resizebox{\linewidth}{!}{\begin{tabular}{|c|c|c|c|}
\hline
\text { Pipeline }           & \text {BNCI2014001}  & \text {BNCI2014004} & \text {Cho2017} \\
\hline \text{CSP+LDA}        & $0.82 \pm 0.17$      & $0.80 \pm 0.15$     & $0.71 \pm 0.15$     \\
\hline \text{MDM}            & $0.82 \pm 0.15$      & $0.78 \pm 0.16$     & $0.63 \pm 0.14$     \\
\hline \text{FgMDM}          & $0.87 \pm 0.12$      & $0.79 \pm 0.15$     & $0.73 \pm 0.13$     \\
\hline \text{TS+EL}          & $0.86 \pm 0.13$      & $0.80 \pm 0.15$     & \textbf{0.76} $\pm$ \textbf{0.14}\\
\hline \text{TS+SVM}         & $0.87 \pm 0.14$      & $0.79 \pm 0.15$     & $0.75 \pm 0.14$    \\
\hline \text{ACM+TS+SVM}     & \textbf{0.92} $\pm$ \textbf{0.10}      & \textbf{0.83} $\pm$ \textbf{0.15}     & $0.74 \pm 0.15$ \\ 
\hline \text{BT-ACM+TS+SVM}  & $0.89 \pm 0.11$      & \textbf{0.83} $\pm$ \textbf{0.14}     & \textbf{0.76} $\pm$ \textbf{0.14}\\ \hline
\hline \text{ShallowConvNet} & $0.86 \pm 0.14$      & $0.72 \pm 0.18$     & $0.74 \pm 0.15$ \\
\hline \text{DeepConvNet}    & $0.82 \pm 0.16$      & $0.72 \pm 0.19$     & $0.72 \pm 0.13$ \\ 
\hline \text{EEGNet}         & $0.77 \pm 0.19$      & $0.70 \pm 0.20$     & $0.67 \pm 0.16$ \\
\hline 
\end{tabular}}
\label{table:rhlh}
\end{table}

\begin{figure*}[ht]  
    \centering
    \centering
     \subfloat[]{%
            \includegraphics[width=0.5\linewidth]{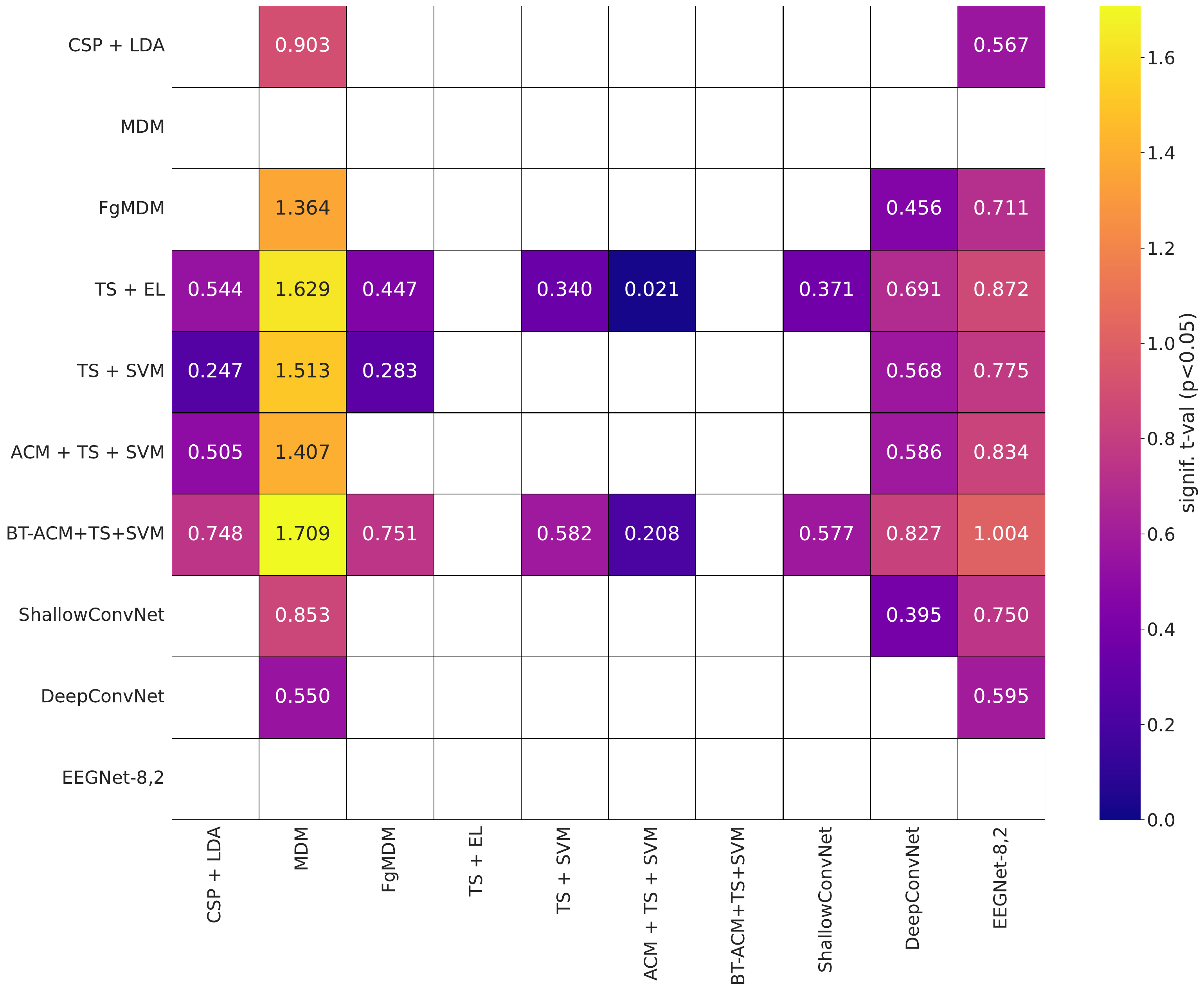}}
            \hfill
   \subfloat[]{%
            \includegraphics[width=0.45\linewidth]{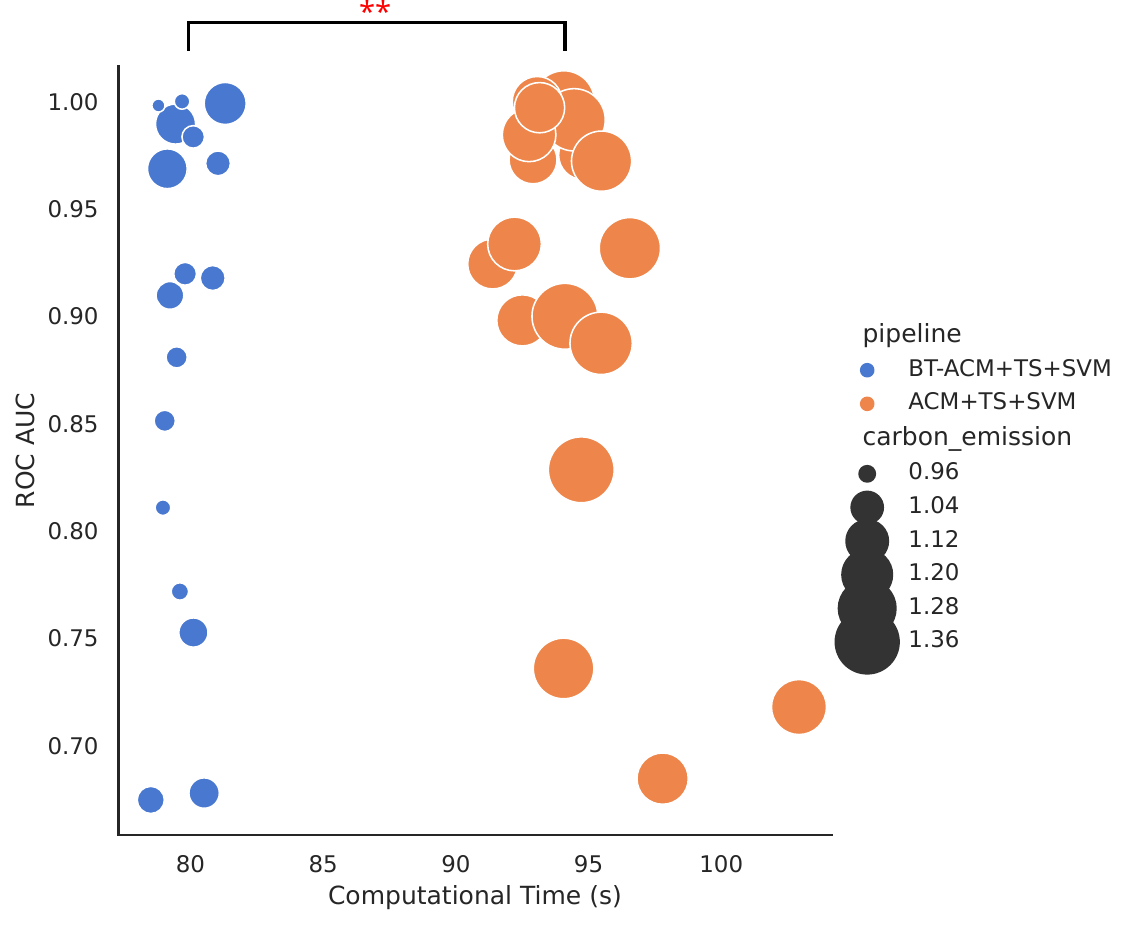}}
            \hfill
    \\
   \subfloat[]{%
            \includegraphics[width=0.30\linewidth]{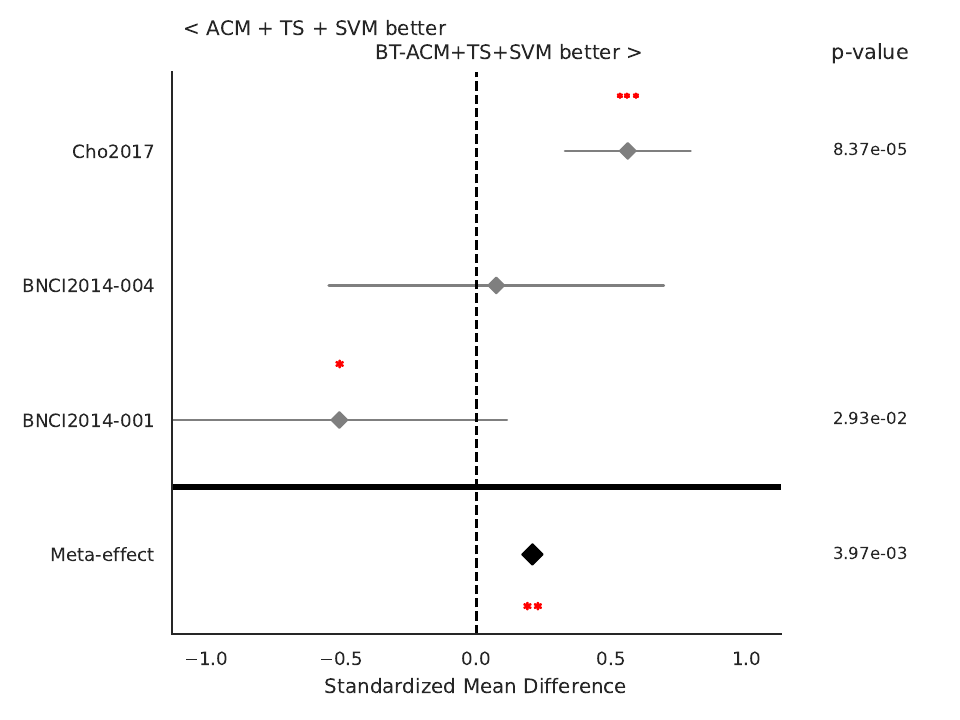}}
            \hfill
   \subfloat[]{%
            \includegraphics[width=0.30\linewidth]{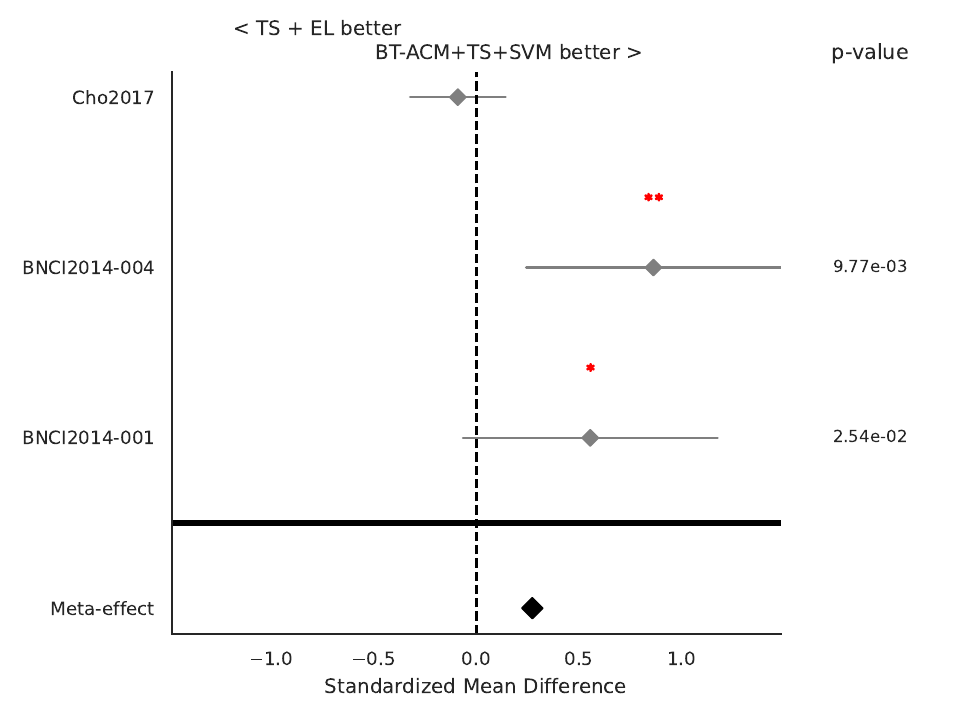}}
            \hfill
   \subfloat[]{%
            \includegraphics[width=0.30\linewidth]{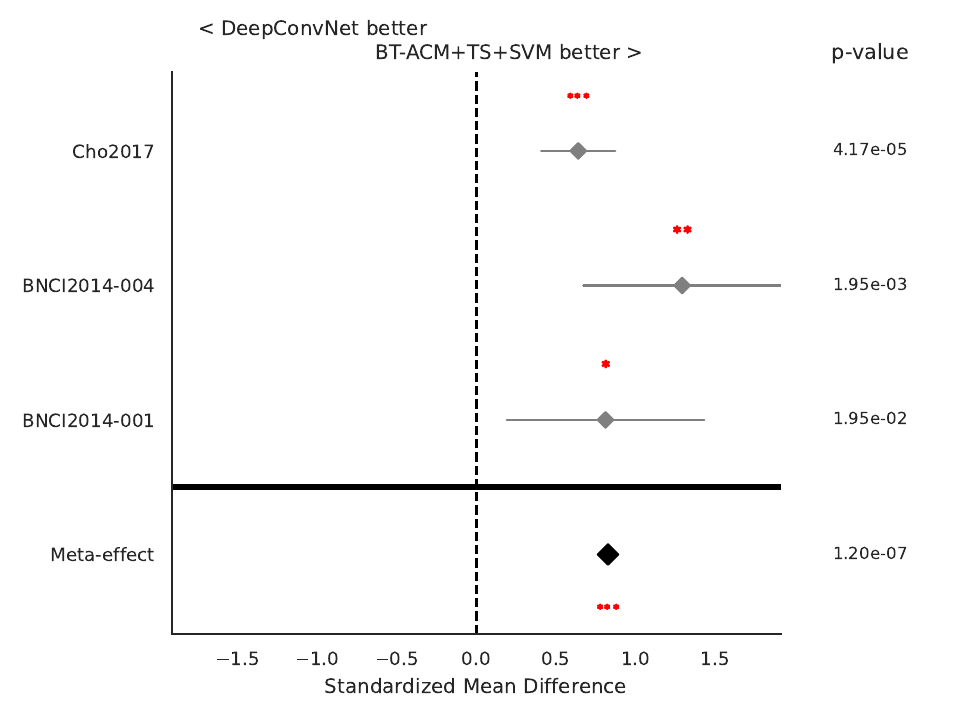}}
            \hfill

    \caption{Results for Right vs Left hand classification, using WS evaluation. Plot (a) provides a combined 
 meta analysis (over all datasets) of the different pipelines. It shows the significance that the algorithm on the y-axis is better than the one on the x-axis. The color represents the significance level of the difference of accuracy, in terms of t-values.  We only show significant interactions ($p < 0.05$). Plots (b) summarizes the computational time and carbon footprint of ACM+TS+SVM vs BT-ACM+TS+SBM. Plots (c), (d) and (e) show the meta analysis of BT-ACM+TS+SVM against respectively ACM+TS+SVM (Grid), TS+EN, DeepConvNet. We show the standardized mean differences of p-values computed as one-tailed Wilcoxon signed-rank test for the hypothesis given as title of the plot. The gray bar denotes the $95\%$ interval. * stands for $p < 0.05$, ** for $p < 0.01$, and *** for $p < 0.001$.
    }
    \label{fig:rhlh}
\end{figure*}

The various approaches of this study are compared in Tab.~\ref{table:rhlh}. A detailed picture of the results and the statistical significance of these results is provided in Fig.~\ref{fig:rhlh}.

Overall, our method scores the best in 2 datasets - BNCI2014001 and Cho2017 - and obtains similar performance with respect to ACM+TS+SVM and TS+EL respectively for the third dataset. Across other datasets evaluated, our approach consistently delivers results that closely rival those of other leading algorithms, with a marginal performance deviation of no more than $1\%$ in the AUC score with the only exception of BNCI2014001 dataset (where only the more costly ACM+TS+SVM does better).

Moreover, a comprehensive analysis across multiple datasets underscores a statistically significant performance enhancement achieved by BT-ACM+TS+SVM compared to all considered algorithms, with the sole exception of TS+EL, where the outcomes are remarkably similar (Fig.~\ref{fig:rhlh} (a) and (d)). This evidence collectively affirms the superiority of our method, not only in achieving high-performance benchmarks, but also in maintaining competitive results across different datasets.

We further explored the carbon emission and the computational time. To perform this analysis, we have run all the algorithms on the same hardware, a Dell C6420 dual-Xeon Cascade Lake SP Gold 6240 @ 2.60GHz. Furthermore, in order to conduct a fair comparison, we considered ACM+TS+SVM and BT-ACM+TS+SVM with the same number of parameter to optimize, i.e. we optimize the order and the lag of the augmentation procedure in the range $[1-10]$ without any optimization of the SVM parameter. Fig.~\ref{fig:rhlh} (b) shows the timings and estimated carbon footprint for both the ACM+TS+SVM and the BT-ACM+TZ+SVM pipelines over the dataset BNCI2014001 composed by 9 subject over 2 sessions. The carbon footprint was estimated using Code Carbon~\cite{benoit-victor-etal:24} and expressed as gCO2 equivalent emission. 


\mysection{Discussion}

Our analysis across multiple datasets demonstrates (see Fig.~\ref{fig:rhlh} (a)) that the BT-ACM+TS+SVM algorithm not only competes with but frequently surpasses the performance of current state-of-the-art methodologies. The only exception to this trend is when compared to the TS+EL algorithm, where our results are statistically indistinguishable (Fig.~\ref{fig:rhlh} (d)). Despite their close relationship, the BT-ACM+TS+SVM algorithm shows a significant (even if small) superior performance compared to the ACM+TS+SVM algorithm (Fig.~\ref{fig:rhlh} (c)). 

We also noticed an improved stability over changes of the SVM parameter, which thus does not need to be optimized as we did in the TS+SVM and ACM+TS+SVM cases (see Table~\ref{table:Pipeline}).

In addition to its improved classification performance, the BT-ACM+TS+SVM algorithm also exhibits significant advancements in computational efficiency. The difference in computational time is statistically significant as we pass from a mean time of $(94.58 \pm 2.62)s$ for the ACM methodology to $(79.71 \pm 0.80)s$ for the BT-ACM+TS+SVM approach. The results have the same level of statistical significance for the carbon emission. Note that the comparison is done using the same number of parameters. The fact that we did not optimize the SVM regularization parameter is not the reason for this improvement.

It is also noteworthy that the ACM methodology exhibits a significant variability in computational times across different sessions and subjects, indicating a fluctuation in performance consistency. In contrast, the BT-ACM approach demonstrates enhanced stability, showcasing a more uniform and predictable computational time.


\mysection{Conclusion}

Throughout this research, we focused on the uses of the Block-Toeplitz Augmented Covariance matrix (BT-ACM) for Motor Imagery classification. This methodology extends the current ACM by comprehensively utilizing the Block-Toeplitz properties of the BT-ACM matrix. This approach transforms the classification challenge, enabling a distinct analysis within the separate domain of SPD and Siegel Disk matrices.

This procedure achieves performance generally superior to the state-of-the-art, or at worst comparable. But, the improvement over the standard ACM is not only in terms of ROC-AUC but also in terms of significant reductions in computational costs and carbon emissions.

An interesting future exploration emerges from the fact that we are not using directly the property of the blocks contained in the BT-ACM matrix since the metric does not belong to any known manifold. In order to treat the problem, we were forced to use the Verblusky coefficient transform that introduces possible errors and complications. This means that it might be interesting to develop the mathematical framework for the direct treatment of such BT matrices.

\mysection{Acknowledgment}

This work has been partially financed by a EUR DS4H/Neuromod fellowship. The authors are grateful to the OPAL infrastructure from Université Côte d'Azur for providing resources and support. 

\mysection*{Data and Code Availability}

The codes used to produce the results of this study are publicly available in this Github repository: 
\url{https://github.com/carraraig/BTACM_BCI}. 

\printbibliography






\end{document}